\documentclass[11pt,a4paper]{article}
\usepackage[T1]{fontenc}
\usepackage[latin1]{inputenc}
\usepackage{amsmath,amssymb,amsthm}
\def\be{\begin{equation}}
\def\ee{\end{equation}}
\def\bea{\begin{eqnarray}}
\def\eea{\end{eqnarray}}

\newcommand{\rmd}{\mathrm{d}}
\newcommand{\rme}{\mathrm{e}}

\newcommand{\ket}[1]{|\kern.3ex#1\kern.3ex\rangle}
\newcommand{\bra}[1]{\langle\kern.3ex #1 \kern.3ex|}
\newcommand{\mean}[1]{\langle#1 \rangle} 

\begin{document}

\begin{center}
{\Large \bf  Random Aharonov-Bohm vortices and some exact
families of integrals: Part II}\\[0.5cm]

{\large \bf Stefan Mashkevich}\footnote{mash@mashke.org}\\[0.1cm]
Schr\"odinger, 120 West 45th St., New York, NY 10036, USA \\ and \\
Bogolyubov Insitute for Theoretical Physics, 03143 Kiev, Ukraine\\[0.4cm]
{\large \bf St\'ephane Ouvry}\footnote{ouvry@lptms.u-psud.fr}\\[0.1cm]
Universit\'e Paris-Sud, Laboratoire de Physique Th\'eorique et Mod\`eles
Statistiques\footnote{Unit\'e Mixte de Recherche CNRS-Paris Sud, UMR 8626}\\
91405 Orsay, France
\\[0.2cm]
\end{center}

\begin{abstract}
At 6th order in perturbation theory,
the random magnetic impurity problem
at  second order in impurity density
narrows down to the evaluation of a single Feynman diagram
with maximal impurity line crossing. This diagram can be  rewritten as  a sum of
ordinary integrals and  nested double integrals of products of the modified Bessel
functions $K_{\nu}$ and $I_{\nu}$, with $\nu=0,1$.
That sum, in turn, is shown to be a linear
combination with rational coefficients of $(2^5-1)\zeta(5)$,
$\int_0^{\infty}u \,K_0(u)^6\,\rmd u$ and $\int_0^{\infty}u^3 K_0(u)^6\,\rmd u$.
Unlike what happens at lower orders, these
two integrals are not linear combinations with
rational coefficients of Euler sums, even though they appear in combination with $\zeta(5)$.
On the other hand, any integral
$\int_0^{\infty}u^{n+1} K_0(u)^p (uK_1(u))^q \,\rmd u$ with  weight $p+q=6$
and an even $n$ is shown to be a linear combination with rational
coefficients of the above two integrals and $1$,
a result that can be easily generalized to any weight $p+q=k$.
A matrix recurrence relation in $n$ is built for such integrals.
The initial conditions are such that the asymptotic behavior is determined by the smallest
eigenvalue of the transition matrix.
\end{abstract}

\section{Introduction}

In Ref.~\cite{nous}, the quantum problem of a charged particle
in a plane, coupled to a random Poissonian
distribution  of infinitely thin impenetrable
Aharonov-Bohm flux tubes (magnetic vortices) perpendicular to the plane was considered.
The relevant parameters are $\alpha = \phi/\phi_0$,
where $\phi$ is the flux of a tube and $\phi_0$ the flux
quantum, and the mean impurity density $\rho=N/V$,
where $N$ is the number of impurities and $V$ the area
(in the thermodynamic limit, $N, V\to\infty$ with $N/V$ fixed).
Periodicity $\alpha\to \alpha+1$
and symmetry with respect to $\alpha=1/2$ imply that the $N$-impurity partition
function $Z_N$ is invariant under $\alpha\to 1-\alpha$, and depends only on
$\alpha(1-\alpha)$.

One was interested in the average partition function 
\be \mean Z = \rme^{-\rho V}\sum_N{(\rho V)^N\over N!}\mean{Z_N}\ee
i.e.,
\bea\label{part} {\mean Z\over Z_0} & = & 1+
\rho V\left[{\mean{Z_1}\over Z_0}-1\right]+
\frac{(\rho V)^2}{2!}\left[{\mean{Z_2}\over Z_0}
-2{\mean{Z_1}\over Z_0}+1\right] \nonumber \\ 
& & {}+
\frac{(\rho V)^3}{3!}\left[{\mean{Z_3}\over Z_0} - 3{\mean{Z_2}\over Z_0}
+3{\mean{Z_1}\over Z_0}-1\right]+ \ldots\eea
where $Z_0$ is the free partition function.

With account for the $\alpha$ dependence of $Z_N$,
Eq.~(\ref{part}) becomes a sum of
terms proportional to $\rho^n\alpha^m$ with $m \ge n$.
For small $\alpha$, the leading terms are $(\rho\alpha)^n$.
They yield the partition function of the charge in the mean magnetic field $\rho\alpha\phi_0$,
which replaces the local magnetic field $\phi\sum_{i=1}^N\delta(\vec r-\vec r_i)$,
$\vec r_i$ being the location of the $i$-th impurity.
Terms with $m>n$ are perturbative corrections to the mean-field expansion,
which originate from disorder effects.
For the 1-impurity case \cite{Comtet}, which is exactly solvable,
one obtains $\mean{Z_1} - Z_0=\alpha(\alpha-1)/2$.
For the 2-impurity case, nontrivial Feynman diagrams with maximal
impurity line crossing appear at order
$\rho^2\alpha^4$, i.e., an electron interacting 4 times with 2 impurities;
at order $\rho^2\alpha^6$, i.e., an electron interacting 6 times with 2 impurities, etc.
Knowing the $\rho^2\alpha^4$ Feynman diagram is sufficient to get a rather precise
analytical estimate of the critical disorder coupling constant
$\alpha_\mathrm{c} \simeq 0.35$, above which oscillations in the density of states,
corresponding to the Landau levels in the mean magnetic field picture, disappear.
That is a clear indication that the system becomes fully disordered.
Note, on the other hand, that at weak disorder, when $\alpha\le\alpha_\mathrm{c}$,
the broadening of the Landau levels due to disorder
fits nicely into the Integer Quantum Hall Effect paradigm.

In Ref.~\cite{nous}, the $\rho^2\alpha^4$  diagram was reduced to
a multiple temperature integral

\be I_{\rho^2\alpha^4}=\int_0^{\beta}\rmd\beta_1\int_0^{\beta_1}\rmd\beta_2
\int_0^{\beta_2}\rmd\beta_3\int_0^{\beta_3}\rmd\beta_4\bigg(
{2\over \beta}-{(a+c)(b+d)\over abc +bcd+cda+dab}\bigg)
\label{ir2a4}\ee 
where $a=\beta_1-\beta_2$, $b=\beta_2-\beta_3$, $c=\beta_3-\beta_4$, $d=\beta_4-\beta_1+\beta$.
A direct step-by-step integration  gave   
\be\label{well} I_{\rho^2\alpha^4}=
\beta^3 \left({1\over 48}-{\tilde\zeta(3)\over 16} \right)\ee  
that is, a linear combination  of $1$ and  $\tilde\zeta(3)=7\zeta(3)/2$
with rational coefficients.
One inferred by connecting the integral (\ref{well})
to the partition function that 
\be \frac{\mean{Z_2}}{Z_0}
-2\frac{\mean{Z_1}}{ Z_0}+ {1}=\frac{1}{Z_0^2}\left[{1\over 6}\alpha^2+
\left({1\over 24}-{\tilde\zeta(3)\over 8}\right)\alpha^4+\ldots\right]\ee 

Thus, an Euler sum of level $3$ emerges, which fits into the general
scheme of Feynman diagram expansion in perturbative field theory \cite{Kreimer},
where Euler sums are known to play a central role.
These sums are defined as
\[\zeta(n_1,n_2,\ldots,n_p)=
\sum_{k_i>k_{i+1}\ge 1}\prod_{i=1}^p{(\pm)^{k_i}\over k_i^{n_i}}\]
where $n=n_1+n_2+\ldots+n_p$ is the level of the sum. At level $n$,
the simplest sums are $\zeta(n)=\sum_{k=1}^{\infty}{1/ k^n}$
and $\zeta_{a}(n)=\sum_{k=1}^{\infty}{(-1)^k/ k^n}$, with
${(2^n-1)}\zeta(n)/2^n={(\zeta(n)-\zeta_a(n))/2}$. 

On the other hand, Eqs.~(\ref{ir2a4})--(\ref{well}) imply that
the nontrivial part of $I_{\rho^2\alpha^4}$ is
 \[\int_0^{\beta}\rmd\beta_1\int_0^{\beta_1}\rmd\beta_2
\int_0^{\beta_2}\rmd\beta_3\int_0^{\beta_3}\rmd\beta_4 \,
{(a+c)(b+d)\over abc +bcd+cda+dab}=\beta^3{1+\tilde\zeta(3)\over 16}\] 
In Ref.~\cite{nousbis}, algebraic  manipulations and a Laplace transform
with respect to $\beta$ let one factorize  this multiple integral as 
\[{1\over 2}\int_{a,b,c,d=0}^{\infty} \rmd a\, \rmd b\, \rmd c\, \rmd d\, \int_0^{\infty}
\rmd t\, {1\over cd} \,
\rme^{-(a+b+c+d)-t\left({1\over a}+{1\over b}+{1\over c}+{1\over d}\right)}
= \int_{0}^{\infty}u \, K_0(u)^2 (uK_1(u))^2 \,\rmd u \]
where $K_{\nu}(u)$ are the modified Bessel functions ($t=u^2/2$). 
Consequently,
\be\label{zeta3}
\int_{0}^{\infty}u\, K_0(u)^2 (uK_1(u))^2 \,\rmd u = {1+{\tilde\zeta(3)}\over 16}
\ee

More generally, it was shown in \cite{nousbis}
by means of a simple integration by parts that any integral of the form
\be\label{fafa}\int_0^{\infty}u^{n+1} K_0(u)^p (uK_1(u))^q \,\rmd u\ee
with  weight\footnote{Here and in the sequel we define the weight of an integral
of a product of Bessel functions $K_{\nu}$ as the total power of the $K_{\nu}$'s.}
$p+q=4$ and $n$ even, is a linear combination
with rational coefficients of $\tilde\zeta(3)$ and $1$.
This was achieved via a $2\times 2$ matrix recurrence relation
for the integrals $\int_0^{\infty}u^{n+1} K_0(u)^4 \rmd u$
and $\int_0^{\infty}u^{n+1} K_1(u)^4 \rmd u$, with $n\ge 4$
(see Sec.~3 for a derivation of this recurrence
and its straightforward generalization to any weight $k=p+q$). 
A remarkable result is the fact that the initial condition
\be\label{3}\int_0^{\infty}u^{5} K_0(u)^4 \,\rmd u=
{-27+7\tilde\zeta(3)\over 64}\;,
\quad \int_0^{\infty}u^{5}K_1(u)^4\,\rmd u=
{53-9\tilde\zeta(3)\over 64}\ee
happens to be such that the asymptotic behavior of the recurrence relation
for large $n$ is governed by the smaller of the two eigenvalues
$\{1/16,1/4\}$ of the asymptotic recurrence matrix.
 
In Ref.~\cite{moi}, these considerations were extended  to integrals
involving $K_{\nu}$ as well as $I_{\nu}$ --- specifically, 
the set (\ref{fafa})  where either one of the $K_0$'s
is replaced by $I_0$ or one of the $K_1$'s is replaced by $I_1$
(so one has a product of three $K_{\nu}$'s and one $ I_{\nu}$).
These integrals were shown, again via an elementary integration by parts,
to be linear combinations with rational coefficients of $3\zeta(2)$  and $1$. 
The $2\times 2$ recurrence matrix  is identical to the one obtained 
in the previous case, up to a minus sign in the off-diagonal elements.
The initial condition
\be \int_0^{\infty}u^{5} K_0(u)^3I_0(u)\,\rmd u={21\zeta(2)\over 128}\;,\quad
\int_0^{\infty} u^5 K_1(u)^3I_1(u) \,\rmd u={27\zeta(2)\over 128}\ee
is such that the asymptotic behavior is governed,
as expected, by the bigger of the eigenvalues $\{1/16,1/4\}$
--- it does not for sure coincide with the unique  initial condition (\ref{3})
associated with the smallest eigenvalue.

\section{The $\rho^2\alpha^6$ diagram with maximal impurity line crossing}

Integrals of products of modified Bessel
functions appear to play a central role in
the perturbative analysis  of the 2-impurity problem.
Indeed, whereas the $\rho^2\alpha^5$ diagrams can be easily shown
to reduce to $\rho^2\alpha^4$ diagrams, 
the relevant  $\rho^2\alpha^6$  diagram with maximal impurity line crossing
is much more arduous to compute.
Following the same route as for the $\rho^2\alpha^4$ diagram, 
and again taking a Laplace transform, one has obtained \cite{nouster}
the expression
\bea \label{wellbis}
I_{\rho^2\alpha^6} & = & 8 \int_0^{\infty}\rmd u \,u \,K_0(u)^2(uK_1(u))^2
\int_0^{u}\rmd x\,(xK_1(x))I_1(x)K_0(x)^2 \nonumber \\
&& {} -4\int_0^{\infty}\rmd u \, u K_0(u)(uK_1(u))[(uK_1(u))I_0(u)-uK_0(u)I_1(u)]
\int_u^{\infty}\rmd x\, x\,K_0(x)^2K_1(x)^2 \nonumber\\ 
&& {} +\int_0^{\infty}u\,K_0(u)^4(uK_1(u))^2 \rmd u 
\eea
It contains not only, in analogy with order  $\rho^2\alpha^4$,
the integral  \[\int_0^{\infty}u \,K_0(u)^4(uK_1(u))^2\,\rmd u\] of weight $6$,
but also a particular combination  of  nested double integrals of products
of modified Bessel functions $K_{\nu}$ and $I_{\nu}$.
Components of those double integrals, if
integrated indivually from $0$ to $\infty$,
are, as already demonstrated\footnote{Note, however, that
$\int_u^{\infty}x\,K_0(x)^2 K_1(x)^2\,\rmd x$
would be divergent if one set $u=0$.}, linear combinations
of either $\tilde\zeta(3)$ and $1$ or of $3\zeta(2)$ and $1$.
For example, in addition to Eq.~(\ref{zeta3}), one has
  \[\int_0^{\infty}(uK_1(u))I_1(u)K_0(u)^2\,\rmd u={\zeta(2)\over 8}\]
\[\int_0^{\infty}u\,K_0(u)(uK_1(u))^2I_0(u)\,\rmd u ={8+3\zeta(2)\over 32}\]
\[\int_0^{\infty}u^2K_0(u)^2(uK_1(u))I_1(u)\,\rmd u={8-3\zeta(2)\over 32 }\]

In effect, there is a mapping,  via an integral, of a product of $K_{\nu}$%
---or a product of $K_{\nu}$  and $I_{\nu}$---onto 
a linear combination with rational coefficients
of $\zeta(3)$ and $1$---or $\zeta(2)$ and $1$, respectively:
\[(f)\to \int_0^{\infty}f(u)\rmd u =\zeta[f]\]
For the double integrals, the same scheme is at work,
but now the mapping is to a ``polyzeta'' object:
\[(f,g)\to \int_0^{\infty}f(u)\rmd u \int_0^u g(x)\rmd x=\zeta[f,g]\]
Since \[\int_0^{\infty}f(u)\rmd u \int_0^u g(x)\rmd x =
\int_0^{\infty}f(u)\rmd u \int_0^{\infty}
g(x) \rmd x -\int_0^{\infty}g(u)\rmd u \int_0^u f(x)\rmd x \] 
one has
\[\zeta[f,g]=\zeta[f]\zeta[g]-\zeta[g,f]\]
in analogy with the relation involving the standard polyzeta function
$\zeta(p,q)=\sum_{n>m}{1\over n^p}{1\over m^q}$:
\be\label{poly}\zeta(p,q)=
\zeta(p)\zeta(q)-\zeta(p+q)-\zeta(q,p)\ee

By analogy with lower orders, one might expect the $\rho^2\alpha^6$
corrections to be a linear combination with rational coefficients
of Euler sums up to a certain level.
The structure of the double integral in Eq.~(\ref{wellbis}) clearly indicates
that the highest level should be 5.
Indeed, the constituent single integrals in the first and second terms
reduce to levels 3 and 2, respectively; by virtue of Eq.~(\ref{poly}),
the product $\zeta(3)\zeta(2)$ is associated with $\zeta(5)$.
The last term in (\ref{wellbis}), 
$\int_0^{\infty}uK_0(u)^4(uK_1(u))^2\,\rmd u$, is a Bessel integral of weight 6.
Given that an integral (\ref{zeta3}) of weight 4
is related to $\zeta(3)$, i.e., level 3, this agains hints at level 5
in the case at hand.

However, a search for an integer relation with the PSLQ algorithm \cite{PSLQ}
does not confirm this expectation. On the contrary, it indicates that 
\be\label{PSLQ}
{I_{\rho^2\alpha^6}}={1\over 30}\int_{0}^{\infty}u \, K_0(u)^6\,\rmd u
+ {1\over 20}\int_{0}^{\infty}u^3 K_0(u)^6\,\rmd u -{31\over 160}\,\zeta(5)
\ee
is a linear combination with rational coefficients of not only,
as expected, a level $5$ Euler sum $(2^5-1)\zeta(5)$,
but also of two numbers of weight 6,
$\int_{0}^{\infty}u \, K_0(u)^6\, \rmd u$ and
$\int_{0}^{\infty}u^3 K_0(u)^6\, \rmd u$,
neither of which is a linear combination with rational coefficients of 
Euler sums of level 5.

Moreover, the same  PSLQ search shows that the last term in Eq.~(\ref{wellbis})
is a linear combination of the same two numbers:
\be\label{toto}\int_{0}^{\infty}u \, K_0(u)^4(uK_1(u))^2\,\rmd u =
{2\over 15}\int_{0}^{\infty}u \, K_0(u)^6\,\rmd u
-{1\over 5}\int_{0}^{\infty}u^3 \, K_0(u)^6\,\rmd u \ee  
Hence, also the sum of double integrals
in (\ref{wellbis}) is by itself a linear combination
with rational coefficients of the same three numbers 
that appear in (\ref{PSLQ})---specifically,
$-\int_{0}^{\infty}u \, K_0(u)^6\,\rmd u/10
+\int_{0}^{\infty}u^3\,K_0(u)^6\,\rmd u/4 -31\zeta(5)/160$.

\section{Integrals $\int_0^{\infty}u^{n+1} K_0(u)^p (uK_1(u))^q \, \rmd u$ with even $n$}

Equation (\ref{toto}) for weight $6$, together with the integrals
of weight $4$ in \cite{nousbis, moi}, are but particular cases of a much more general pattern
involving integrals of the form $\int_0^{\infty}u^{n+1} K_0(u)^p (uK_1(u))^q \, \rmd u$
with even $n$. 

At a given weight $k = p+q$, denote
\be\label{injk}
I_{n,j}^{(k)} = \frac{1}{n!}\int_0^{\infty}u^{n+1}K_0(u)^{k-j}K_1(u)^j\rmd u
\ee
where $j=0,1,2,\ldots,k$.
For the integral to be finite, $n\ge j-1$ is required.
Integration by parts, taking into account that
$dK_0(u)/du = -K_1(u)$ and $d(uK_1(u))/du = -uK_0(u)$,
gives a recurrence relation 
\be
\label{rec}I_{n,j}^{(k)}={n+1\over n-j+2}
\left[ jI_{n+1,j-1}^{(k)} + (k-j)I_{n+1,j+1}^{(k)} \right]
\ee

The mapping $\{I_{n+1,j}^{(k)}\} \to \{I_{n,j}^{(k)}\}$
induced by Eq.~(\ref{rec})  and valid\footnote{We choose here,
as a matter of convenience, to start
the recurrence at $n=k$.} for $n\ge k$ involves a tridiagonal matrix 
\be
A_n^{(k)} = \left(
\begin{array}{cccccccc}
0 &  \frac{k(n+1)}{n+2}&0&0&0&\ldots&0&0\\
1& 0& k-1&0&0&\ldots&0&0\\ 
0&  \frac{2(n+1)}{n}&0& \frac{(k-2)(n+1)}{n}&0&\ldots&0&0\\ 
0&  0& \frac{3(n+1)}{n-1}& 0& \frac{(k-3)(n+1)}{n-1}&\ldots& 0&0\\ 
0& 0& 0& \frac{4(n+1)}{n-2}& 0& \ldots& 0&0\\ 
\ldots& \ldots& \ldots& \ldots& \ldots& \ldots& \ldots& \ldots\\
0& 0&0&0&0& \ldots& \frac{k(n+1)}{n-k+2}&0\\ 
\end{array}\right)
\ee
One has
$\det A_n^{(k)}=-{(3)^2(5)^2\ldots(k)^2(n+1)^{2k+1}\over(n+2)(n+1)(n)\ldots(n-k+2) }$
when $k$ is odd, and $0$ when $k$ is even, meaning that in the latter case
the $I_{n,j}^{(k)}$'s are
linearly related. Indeed, for an even $k$, one has
\be\label{link}
\sum_{l=0}^{k/2} (-1)^l (n-2l+2)
{k/2 \choose l} I_{n,2l}^{(k)} = 0
\ee
By applying relation (\ref{rec}) twice, one obtains
a mapping $\{I_{n+2,j}^{(k)}\} \to \{I_{n,j}^{(k)}\}$:
\bea
I^{(k)}_{n,j} & = & \frac{(n+1)(n+2)(j-1)j}{(n-j+2)(n-j+4)}\,I^{(k)}_{n+2,j-2} \nonumber \\
&& + \frac{(n+1)(n+2)}{n-j+2}\left[\frac{(j+1)(k-j)}{n-j+2} +
\frac{j(k-j+1)}{n-j+4}\right]I^{(k)}_{n+2,j} \nonumber \\
&& + \frac{(n+1)(n+2)(k-j-1)(k-j)}{(n-j+2)^2}\,I^{(k)}_{n+2,j+2} \label{rec2}
\eea
This can be inverted, as long as takes into account
the linear relation (\ref{link}) in the case of an even $k$.

The recurrence relations (\ref{rec}),~(\ref{rec2}) conserve the parity of $n-j$;
thus, all $I^{(k)}_{n,j}$'s are divided into two subsets, and the relations
operate separately within each subset.
For reasons that will become clear later, we focus on one of those---the one
for which $n-j$ is even.
To span this subset, it is enough to assume that $n$ is even, then take the set
\be\label{family1}I_{n,0}^{(k)}\,,\quad I_{n,2}^{(k)}\,,\quad \ldots,\quad
I_{n,k}^{(k)}\,,\quad n\ge k\quad (k \mbox{ even})\ee 
[remembering that these integrals are linearly related via Eq.~(\ref{link})], or
\be\label{family2}I_{n,0}^{(k)}\,,\quad I_{n,2}^{(k)}\,,\quad \ldots,\quad
I_{n,k-1}^{(k)}\,,\quad n\ge k-1\quad (k \mbox{ odd})\ee
and  note that the integrals
\be\label{fa1} I_{n+1,1}^{(k)}\,,\quad I_{n+1,3}^{(k)}\,,\quad \ldots, \quad I_{n+1,k-1}^{(k)}
\quad (k \mbox{ even})\ee
and
\be\label{fa2} I_{n+1,1}^{(k)}\,,\quad I_{n+1,3}^{(k)}\,, \quad \ldots, \quad I_{n+1,k}^{(k)}
\quad (k \mbox{ odd})\ee
are related to (\ref{family1}) and (\ref{family2}), respectively,
by inverting the recurrence (\ref{rec})
[again, taking into account (\ref{link}) if $k$ even].
The union of sets (\ref{family1})--(\ref{fa2}), at a given $k$, is
tantamount to the family  of integrals 
\be\label{set}\int_0^{\infty}u^{n+1} K_0(u)^p (uK_1(u))^q \, \rmd u\ee
with  weight $p+q=k$ and an even $n$.

For an even $k$, by virtue of Eq.~(\ref{rec2}),
all integrals from the family (\ref{family1}),~(\ref{fa1})
can be expressed as a linear combination with rational coefficients of
 the initial conditions
\be\label{ouf1}I_{k,0}^{(k)},\quad I_{k,2}^{(k)},\quad \ldots,\quad I_{k,k}^{(k)}\ee
still taking into account Eq.~(\ref{link}).
Noting further that Eq.~(\ref{rec}) implies
$2I_{k-1,k-1}^{(k)}={k}(I_{k,k}^{(k)}+(k-1) I_{k,k-2}^{(k)})$
and that, trivially, $I_{k-1,k-1}^{(k)}=1/k!$,
one can trade $I_{k,k}^{(k)}$ for $1$.
Using (\ref{link}), one can get rid of one more  element:
thus, all integrals (\ref{family1}),~(\ref{fa1}) are linear combinations
with rational coefficients of a basis made of the $k/2$ independent
(in the sense that none of them is a linear combination with rational coefficients
of the others) numbers
\be\label{bkeven}
\{1\,, I_{k,0}^{(k)}\,,  I_{k,2}^{(k)}\,, \ldots,  I_{k,k-4}^{(k)}\}
\ee

Likewise, for $k$ odd, starting with the initial conditions
\be\label{ouf2}I_{k-1,0}^{(k)},\quad I_{k-1,2}^{(k)}, \quad\ldots, \quad I_{k-1,k-1}^{(k)}\ee
and again using $I_{k-1,k-1}^{(k)}=1/k!$,
one concludes that any integral in (\ref{family2}),~(\ref{fa2})
is a linear combination with rational coefficients of
the basis made of the $(k+1)/2$ independent numbers
\be\label{bkodd}
\{1\,,I_{k-1,0}^{(k)}\,, I_{k-1,2}^{(k)}\,, \ldots, I_{k-1,k-3}^{(k)}\}
\ee

Last but not least, one can also show, by applying (\ref{rec}) appropriately
for $0\le n\le k$, that for $k$ even, the basis (\ref{bkeven}) can be mapped on  the basis
\be\label{ba1} \{1\,, I_{0,0}^{(k)}\,, I_{2,0}^{(k)}\,,
 \ldots, I_{k-4,0}^{(k)}\}\ee
and for $k$ odd, the basis (\ref{bkodd})
can mapped on
\be\label{ba2}\{ 1\,, I_{0,0}^{(k)}\,, I_{2,0}^{(k)}\,,
 \ldots, I_{k-3,0}^{(k)}\}\ee

Therefore, any integral in the set (\ref{set})  is a linear combination
with rational coefficients of the basis (\ref{ba1}) or (\ref{ba2}),
for $k$ even or odd, respectively.

Consider now the asymptotic regime, $n\to\infty$.
In that limit, Eq.~(\ref{rec2}) becomes
\be
I^{(k)}_{n,j} = A^{(k)}_{j,j-2}I^{(k)}_{n+2,j-2} + A^{(k)}_{j,j}I^{(k)}_{n+2,j}
+ A^{(k)}_{j,j+2}I^{(k)}_{n+2,j+2} 
\ee
where the elements of the tridiagonal transition matrix $A^{(k)}$
(indexed so that their subscripts are always even) no longer depend on $n$:
\be
A^{(k)}_{j,j-2} = (j-1)j\;, \quad
A^{(k)}_{j,j} = 2j(k-j)+k\;, \quad
A^{(k)}_{j,j+2} = (k-j-1)(k-j)
\ee
The $[(k - k\bmod2)/2 + 1]$ eigenvalues of this matrix are:
$k^2$, $(k-2)^2$, $(k-4)^2$, \ldots\,;
the last one is 1 for odd $k$ and 0 for even $k$.
In the latter case, before
inverting the matrix, one has to reduce its dimension by one, using Eq.~(\ref{link}).
Thereupon, the largest eigenvalue of the inverse matrix is 1 or 1/4 for
$k$ odd or even, respectively, whereas the smallest one is $1/k^2$.
Remarkably, as we discovered experimentally, with the
initial conditions (\ref{ouf1}),~(\ref{ouf2})  it is the
smallest eigenvalue that determines the asymptotic behavior of $I^{(k)}_{n,j}$.
In the asymptotic regime itself, this can be understood by noting that
the eigenvector corresponding to said eigenvalue,
i.e., to the $k^2$ eigenvalue of $A^{(k)}$, is $\{1,1,\ldots,1\}$ 
--- because $A^{(k)}_{j,j-2} + A^{(k)}_{j,j} + A^{(k)}_{j,j+2} = k^2$.
Now, in the limit $n\to\infty$, the integral $I^{(k)}_{n,j}$
does not depend on $j$, because the integrand $u^{n+1}K_0(u)^{k-j}K_1(u)^j$
peaks at large values of $u$, where both $K_0(u)$ and $K_1(u)$
can be approximated by their common asymptotic behavior,
$K_{\nu}(u)
\raisebox{-0.53em}{$\stackrel{\textstyle\longrightarrow}{\scriptstyle u\to\infty}$}
\sqrt{{\pi\over 2u}}e^{-u}$.
Therefore, the vector $\{I_{n,j}^{(k)}\}$ becomes, in the asymptotic limit,
proportional to $\{1,1,\ldots,1\}$.
The fact that the initial condition (\ref{ouf1}),~(\ref{ouf2})
leads to this special asymptotic behavior, as well as
the fact that said condition can be expressed
in terms of the basis (\ref{ba1})--(\ref{ba2}),
means that the building blocks of (\ref{ba1})--(\ref{ba2}),
namely  Bessel function integrals $\int_0^{\infty}u^{n+1}\, K_0(u)^{k}\,\rmd u$
with $n$ even, may play some special role in number theory---like Euler sums do
(but again, these integrals are not rational linear combinations of those sums).

As an example, we detail the recurrence relations for weight $k=3$,
where one has computed \cite{nousbis} 
\be\label{ouf}I_{0,0}^{(3)}={1\over 6}\sum_{k=0}^{\infty}{1\over (k+1/3)^2}-{2\over 3}\zeta(2)\ee
One has, for an even $n\ge 2$,
\be\label{reccurence}
\left(
\begin{array}{c}
\displaystyle I_{n,0}^{(3)}\\ \\
\displaystyle I_{n,2}^{(3)}\end{array}\right)
=
\left(
\begin{array}{cc}
\displaystyle \frac{3(n+1)}{n+2} & \displaystyle \frac{6(n+1)}{n+2}\\
\displaystyle \frac{2(n+1)}{n} & \displaystyle \frac{\rule{0em}{1.5em}(n+1)(7n+6)}{n^2} 
\end{array}\right)
\left(
\begin{array}{c}
\displaystyle I_{n+2,0}^{(3)}\\ \\
\displaystyle I_{n+2,2}^{(3)}\end{array}\right)
\ee
The inverse relation is
\be
\left(
\begin{array}{c}
\displaystyle I_{n+2,0}^{(3)}\\ \\
\displaystyle I_{n+2,2}^{(3)}\end{array}\right)
=
\left(
\begin{array}{cc}
\displaystyle \frac{7n+6}{9(n+1)} & \displaystyle -\frac{2n^2}{3(n+1)(n+2)}\\
\displaystyle -\frac{2n}{9(n+1)} & \displaystyle \frac{n^2\rule{0em}{1.5em}}{3(n+1)(n+2)} 
\end{array}\right)
\left(
\begin{array}{c}
\displaystyle I_{n,0}^{(3)}\\ \\
\displaystyle I_{n,2}^{(3)}\end{array}\right)
\ee
which in the asymptotic limit turns into
\be\label{asymp}
\left(
\begin{array}{c}
\displaystyle I_{n+2,0}^{(3)}\\ \\
\displaystyle I_{n+2,2}^{(3)}\end{array}\right)
=\left(
\begin{array}{cc}
\displaystyle  \frac{7}{9} & \displaystyle -\frac{2}{3} \\
\displaystyle  -\frac{2}{9} & \displaystyle \quad \frac{1\rule{0em}{1.5em}}{3}
\end{array}
\right)
\left(
\begin{array}{c}
\displaystyle I_{n,0}^{(3)}\\ \\
\displaystyle I_{n,2}^{(3)}\end{array}\right)
\ee
with eigenvalues $\{1/9,1\}$.

Clearly, as alluded to above, instead of the set (\ref{set}), one could have focused on the set 
$\int_0^{\infty}u^{n+1} K_0(u)^p (uK_1(u))^q\,\rmd u$
with weight $p+q=k$  and $n$ odd---corresponding to the subset of
integrals (\ref{injk}) with odd $n-j$.
The recurrence still operates within this set,
and the same kind of algebra as above is at work.
However, the integrals in this set 
\begin{itemize}
\item
do not   play any role
in the perturbative analysis  of the random magnetic impurity problem
(at least up to $6$th order);
\item
but still lead to an asymptotic behavior
governed by the smallest  eigenvalue of the corresponding asymptotic recurrence matrix.
This can be easily seen in the case $k=3$, where the initial conditions $I_{2,3}^{(3)},\; I_{2,1}^{(3)}$
for the integrals $I_{n,3}^{(3)}$ and $I_{n,1}^{(3)}$ with $n$ even
and the corresponding recurrence relation  lead to an asymptotic governed
by the smallest of the eigenvalues  $\{1/9,1\}$ of the asymptotic matrix.
\end{itemize}
Note finally that at weight $k=n+1$, by  singling out the simplest integral
$I_{0,0}^{(n+1)}$ in the basis (\ref{ba1}) or (\ref{ba2}), 
 one arrives at  the number   
\be
\label{end} \kappa(n)={1\over (n+1)!}\int_0^{\infty}u\, K_0(u)^{n+1}\,\rmd u
\ee
with $\kappa(0)=1$; $\kappa(1)=1/4$; $\kappa(2) = I_{0,0}^{(3)}/3!$,
see Eq.~(\ref{ouf});
$\kappa(3)=7\zeta(3)/(8\times 4!)$, etc. 
As already said, this number 
is analogous to Euler sums of level $n$, but it is not
a rational linear combination of those.
When $n\to \infty$, the ${1/(n+1)!}$ normalization in (\ref{end})
is such that\footnote{This can be easily shown by recognizing that when
$n\to\infty$, the integrand $u\,K_0(u)^{n+1}$ has a peak near the origin, where $K_0(u)$
can be approximated by $K_0(u)\simeq -\log\frac{u}{2\exp[\Gamma'(1)]}$.
It follows that the main contribution to the integral is
$\int_0^{\infty}u\,K_0(u)^{n+1}\,\rmd u \simeq \int_0^{2\exp[\Gamma'(1)]}
u\,(-\log\frac{u}{2\exp[\Gamma'(1)]})^{n+1}\rmd u$, which trivially yields (\ref{endbis}).}
\be
\label{endbis} \rme^{-2\Gamma'(1)}\lim_{n\to\infty}\kappa(n)=\lim_{n\to\infty}(\zeta(n)-1)
\ee
When $n\to -1$, on the other hand,
$\kappa(n)={1/ (1+n)^2}$  plus logarithmic subleading terms and a constant.
Clearly, the function $\kappa(n)$ defined for $n\ge -1$ real
can be analytically continued to the function $\kappa(s)$
defined on the whole complex half-plane $\Re(s)\ge -1$.
 
\section{Conclusion}

We have demonstrated that a  class of integrals involving Bessel
functions, which arise in perturbation theory---in particular, in the
two-dimensional problem of random magnetic impurities---can be expressed,
via recurrence relations,
as linear combinations with rational coefficients of ``basis'' integrals.
Some of the latter, in turn, reduce to Euler sums, but most do not.
Additionally, these same basis integrals, for an even power of the argument,
turn out to generate the unique initial conditions for the
recurrence relations which make the smallest eigenvalue of the transition
matrix determine the asymptotic behavior.
The nature of this phenomenon has yet to be understood more deeply.

It is not only single but also some double nested integrals
that happen to be linear combinations with rational coefficients
of said basis integrals: one that does is the integral
that figures in the 6th order of perturbation theory in the
physical problem at hand.
Understanding what other double integrals fall into the same class
remains another open question.

{\bf Acknowledgements}: One of us (S.O.) would like to thank Jean Desbois
for discussions and some technical help, in particlular at the end of Section 3.

{\bf Note added}: After completion of this work, we became aware of Ref.~\cite{Kreimerbis}
and  in particular of Ref.~26 therein, where some results overlap with those of section 3.

\end{document}